# INDOOR-ATMOSPHERIC RADON-RELATED RADIOACTIVITY AFFECTED BY A CHANGE OF VENTILATION STRATEGY

TSUNEO KOBAYASHI

*Department of Physics, Fukushima Medical University School of Medicine, Fukushima, 960-1295, Japan*

**Abstract:** The present author has kept observation for concentrations of atmospheric radon, radon progeny and thoron progeny for several years at the campus of Fukushima Medical University.  Accidentally, in the midst of an observation term, i.e., February 2005, the facility management group of the university changed a strategy for the manner of ventilation, probably because of a recession: (I) tidy everyday ventilation of 7:30-24:00 into (II) shortened weekday ventilation of 8:00-21:00 with weekend halts.  This change of ventilation manner brought a clear alteration for the concentrations of radon-related natural radioactivity in indoor air.  The present paper concerns an investigation of the effect of the ventilation strategy on the indoor-atmospheric radon-related radioactivity.

Running head:
RADON-RELATED RADIOACTIVITY AFFECTED BY A VENTILATION STRATEGY

Correspondence to: Tsuneo Kobayashi, Department of Physics, Fukushima Medical University, Fukushima City 960-1295, Japan.
Email: tkoba@fmu.ac.jp

INTRODUCTION

Research on natural radioactivity and natural radiation is meaningful for medical students and doctors. Clarification of the behavior of radon-related radioactivity in air is especially an issue of wide importance to radiation protection [1], earthquake prediction [2], etc. Radon ($^{222}$Rn, half life 3.82 d) and thoron ($^{220}$Rn, half life 55 s) are invisible odorless gaseous natural radioactive elements that belong in uranium series and thorium series, respectively [1]. World average of annual effective dose to a human by natural radiation [3] is 2.4 mSv and about half of which is due to internal exposure to progenies of $^{222}$Rn and $^{220}$Rn. Sources of radon and thoron in indoor air are said to be earth and rock beneath the building, building materials, etc.

The present author has kept measurements of natural radiation and/or natural radioactivity especially radon-related radioactivity in air in a room of the author's office at the campus of Fukushima Medical University [4-7]. A change in the manner of ventilation occurred in February 2002, probably because of a recession. This change brought a considerable change for the concentrations of radon-related radioactivity.

This paper investigates the effect of the change of ventilation strategy on the concentrations of atmospheric radon, radon progeny and thoron progeny.

MATERIALS AND METHOD

Pylon TEL (active static-collecting type of Lucas cell, 0.5 L/min) detected radon concentration, and Pylon WLx (spectroscopic method for continuously filtering with a membrane filter, 0.5 L/min) detected radon and thoron progeny concentrations [8]. Both monitors were set to output concentrations every one hour.

Measurement was performed in a room of the author's office (named 2nd Lab). The size of the room was 3.75 m * 6.5 m * 2.7 m = 66 m$^3$. The window of the room was never opened and the doors were shut to the utmost during the measurement. The ventilation system of the university introduces outdoor air through a pre-filter and a main filter made of polyester and moda-acrylic fiber, respectively. The ventilation rate was 130 m$^3$/hour. Because of this strong ventilation system, concentrations of radon-progeny and thoron-progeny in indoor air were found to be less than those in outdoor air [6].

On the one hand, radon concentration never reduces with any filters because radon is an inert gas. In fact, recent preliminary study has shown about 2.8 times higher concentration of radon in indoor air than that in outdoor air. On the other hand,

radon- and thoron-progeny concentration reduces easily when they pass through some kind of filters because progeny are not inert gas but metal elements such as $^{218}$Po (RaA), $^{214}$Bi (RaB) and $^{214}$Pb (RaC). Thus, progeny concentrations are anomalously low under the strong ventilation system.

To see the effect of the change of ventilation system, the data from February 2002 to January 2003 were selected. Just in the center of this term, the ventilation strategy changed from (I) tidy everyday ventilation of 7:30-24:00 into (II) shortened weekday ventilation of 8:00-21:00 with weekend halts. Figure 1 shows typical temporal hourly variation of radon concentration for one month, May as an example, picked up from the term (I) 2002-02-01 to 2003-01-31 and the term (II) 2003-02-01 to 2004-01-31, respectively. In both terms, starting the ventilation indicates sudden decrease of radon concentration since the concentration is much lower in outdoor air than in indoor air. In term (II), weekend halts of ventilation resulted in gently varying high concentration lapse seen once in a while in this figure.

Statistical analyses were made using S-PLUS 6.2J [9]. Paired t-test (two sided, significant level 1%) checked the differences between arithmetic means of three concentration sets before and after the change of ventilation strategy. Furthermore, Shapiro and Wilk's normality test verified the statistical distribution of data. Lastly, one-way layout analysis with multiple comparison confirmed seasonal variations.

RESULTS AND DISCUSSION

Table 1 shows basic statistics for radon concentration (Rn), equilibrium equivalent radon progeny concentration (ECRn) and equilibrium equivalent thoron progeny concentration (ECTn) before and after the change in the manner of ventilation. Arithmetic means of the concentrations of radon and its progeny rose slightly after the change of ventilation, whereas that of thoron progeny increased drastically. Each increase for the three concentration pairs all passed paired t-test as shown in the column of p-value in Table 1. Those increases are attributed to the decrease of the introduction of outdoor air after the change of the ventilation strategy.

A log-normal distribution is one of the well-known characteristics of radon and its progeny. To examine the statistical distribution, the author performed the normality test for the logarithm of each data set (zero-data were omitted). As can be seen from the column of LNT in Table 1, log-normality decreased after the change of ventilation manner. This reduction of log-normality is probably attributed to less introduction of outdoor air after the ventilation change, since outdoor air often indicates log-normal distribution [6, 7].

Figure 2 shows the temporal variation of monthly averaged concentration for radon, radon progeny and thoron progeny separately.  Figure 2 again indicates slight increase of radon and its progeny, and drastic increase of thoron progeny.  All these increases are statistically significant as mentioned before.

To see the seasonal variation, data were separated into four seasons, i.e., spring (March to May), summer (June to August), fall (September to November) and winter (December to February).  Since the change of ventilation strategy occurred in February, sequential data were not available for the winter data, therefore the winter data were picked up as February 2002, December 2002 and January 2003.

Essentially the same seasonal dependency was observed for the term (I) and (II), namely the concentration of radon and radon progeny were high in fall, whereas that of thoron progeny was high in summer.

Seasonal variation of radon and its progeny may be affected by the variation of outdoor air.  As a matter of fact, outdoor radon concentration is also high in fall, which is interpreted as follows.  The outdoor air is stable in autumn [7] and the vertical mixing of air is small in this season, especially in the morning of fine weather, resulting the decreased diffusion rate of radon into atmosphere, therefore radon concentration near the ground remains high.  In spring and winter, wind with high speed carries away radon and its progeny, and in summer, high temperature of the earth increases vertical mixing of air considerably.  The author does not know any information concerning seasonal variation of thoron.

Although the existence of strong ventilation system, filters cannot prevent entry of radon into the room because radon is an inert gas.  Thus, seasonal variations of indoor radon and its progeny resemble that of outdoor ones.  However, thoron will not come from outdoor air because of the short half life of thoron and the long route to the room.

Why concentrations of thoron progeny are high in summer, unlike concentrations of radon and its progeny?  The answer is probably as follows.  The origin of atmospheric thoron in the room is considered to be the materials of the room itself, i.e., the floor, wall, ceiling, etc.  The dissipation of thoron from these materials may depend on temperature.  High temperature in summer may probably increase the dissipation rate of thoron.  Ventilation carries out thoron from the room to outside through exhaust pipes, etc.  Thus, the change of the manner of ventilation system strongly affected thoron and its progeny concentration.

## CONCLUSION

The author investigated the effect of the ventilation strategy on the concentrations of atmospheric radon, radon progeny and thoron progeny.  After all, less ventilation increased radon, radon progeny and thoron progeny.  Interesting high concentration of thoron progeny in summer may probably be attributed to the short half life of thoron.

Observation of the change of the radon-related three concentrations before and after the change of ventilation manner gave us a chance to speculate about the tendency and/or origin of radon-related radioactivity.

## REFERENCES

1. International Commission on Radiological Protection.  Protection against radon-222 at home and at work.  Ann ICRP **23** No.2, 1993.  *In*: ICRP publication 65, Pergamon Press, Oxford, 1993.
2. Yasuoka Y , Shinogi M.  Anomaly in atmospheric radon concentration: a possible precursor of the 1995 Kobe, Japan, Earthquake.  Health Phys, **72**: 759–761, 1997.
3. United Nations Scientific Committee on the Effect of Atomic Radiation.  Sources and effects of ionizing radiation.  UNSCEAR 2000 reports to the general assembly, with scientific annexes.
4. Kobayashi T.  Measurement of natural radiation at the campus of Fukushima Medical University. (in Japanese)  Fukushima Med J, **52**: 1–9, 2002.
5. Kobayashi T, Takaku Y.  Measurement of atmospheric radon daughter concentrations at the campus of Fukushima Medical College.  Fukushima J Med Sci, 35: 29–43, 1989.
6. Kobayashi T, Takaku Y.  Intermittent measurement of $^{222}$Rn and $^{220}$Rn progeny in air for four years.  RADIOISOTOPES, **46**: 603–614, 1997.
7. Kobayashi T: Temporal variation of radon progeny ratio in outdoor air.  Health Phys, **83**: 287–292, 2002.
8. George AC.  State-of-the-art instruments for measuring radon/thoron and their progeny in dwellings—a review.  Health Phys, **70**: 451–463, 1996.
9. Mathematical Systems, Inc.  http://www.msi.co.jp/splus/

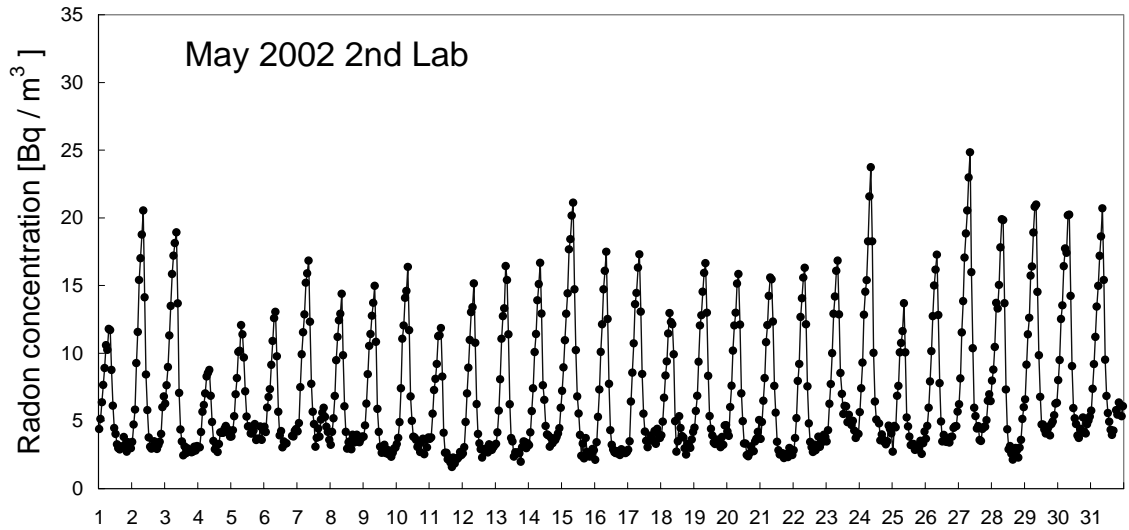

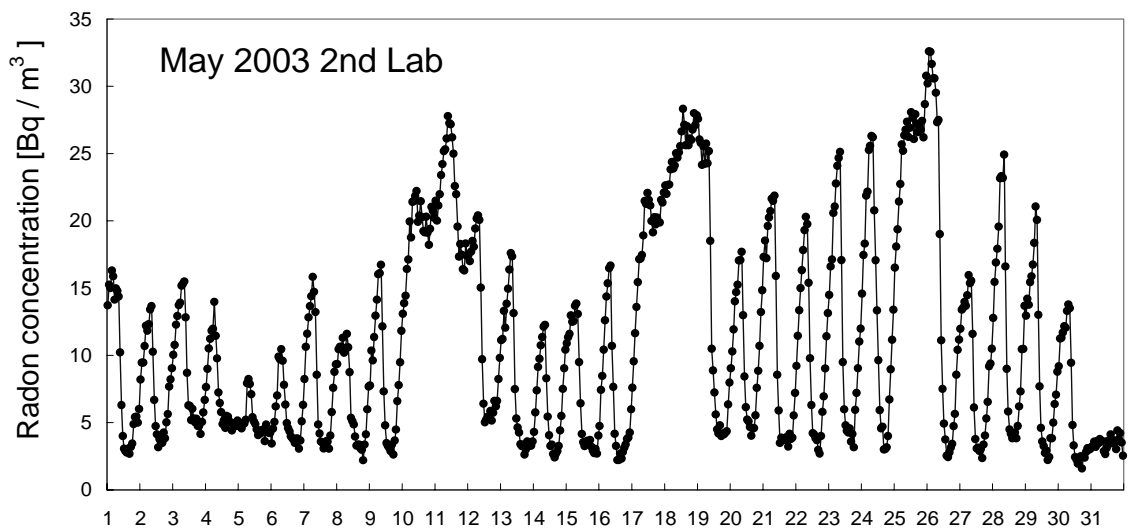

**Figure caption**

Figure 1: Temporal variation of atmospheric radon concentration before (top) and after (bottom) the change of the manner of ventilation system. Closed circles represent hourly concentration. Numbers of 1, 2, 3, … in the abscissa means 5/1, 5/2, 5/3, …, respectively.

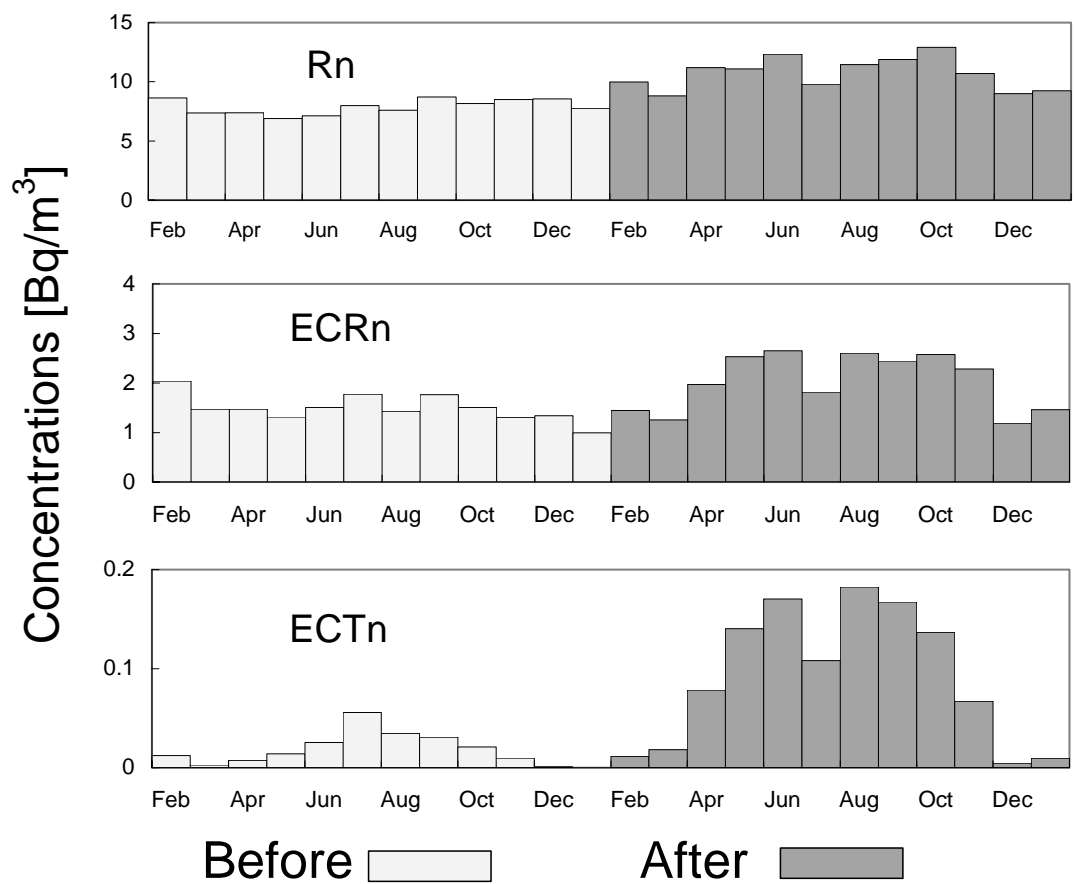

**Figure caption**

Figure 2: Temporal variation of monthly averaged concentrations of atmospheric radon (Rn), radon progeny (ECRn) and thoron progeny (ECTn). Note the scale difference of the ordinates.

Table 1: Basic statistics for concentrations of radon (Rn), equilibrium equivalent concentration of radon progeny (ECRn), and equilibrium equivalent concentration of thoron progeny (ECTn). All values except p-values are in unit of Bq/m$^3$. **Before** and **After** means before and after the change of the manner of ventilation system, respectively. "p-value" denotes probability values from the paired t-test for data sets before and after the change of ventilation (in this column, p-values were all less than $1.0 \times 10^{-10}$). "LNT" means log-normality tests, where ∗∗ and ∗ denotes log-normality with significance level of 1% and 5%, respectively.

|  |  | Median | Arithmetic Mean | p-value | Std. Dev. | Max. | Min. | LNT |
|---|---|---|---|---|---|---|---|---|
| **Before** | Rn | 7.52 | 7.88 |  | 1.78 | 15.65 | 4.12 | ∗∗ |
|  | ECRn | 1.35 | 1.49 |  | 0.77 | 4.48 | 0.17 | ∗ |
|  | ECTn | 0.02 | 0.02 |  | 0.03 | 0.24 | 0 |  |
| **After** | Rn | 9.95 | 10.69 | < 0.01 | 4.36 | 28.44 | 3.10 | ∗ |
|  | ECRn | 1.75 | 2.02 | < 0.01 | 1.25 | 8.33 | 0.08 |  |
|  | ECTn | 0.03 | 0.09 | < 0.01 | 0.11 | 0.50 | 0 |  |